\documentclass[12pt]{article}
\usepackage{hyperref}

\begin{document}

\title{On quantum vs. classical probability}

\author{Jochen Rau\thanks{email: jochen.rau@q-info.org} \\ 
Institut f\"ur Theoretische Physik \\
Johann Wolfgang Goethe-Universit\"at \\
Max-von-Laue-Str. 1, 
60438 Frankfurt am Main,
Germany
}

\date{\today}

\maketitle

\begin{abstract}
Quantum theory shares with classical probability theory many important properties.
I show that this common core regards at least the following six areas,
and I provide details on each of these:
the logic of propositions,
symmetry,
probabilities,
composition of systems,
state preparation
and reductionism.
The essential distinction between classical and quantum theory, on the other hand,
is shown to be joint decidability versus smoothness;
for the latter in particular I supply ample explanation and motivation.
Finally, I argue that beyond quantum theory there are no other generalisations of classical probability theory that are relevant to physics.
\end{abstract}

\newpage

\section{\label{intro}Introduction}

Physical theories which are inherently probabilistic are notoriously difficult to grasp.
Determinism seems so deeply ingrained in our thinking that already classical concepts such as, say, the notion of entropy, maximum-entropy priors or the second law of thermodynamics, albeit in principle well understood \cite{jaynes:papers,balian:book1,balian:book2,rau:physrep}, trigger controversy up to this day \cite{lebowitz:boltzmann,barnumetal:letters}
--- let alone the conceptual basis of quantum theory \cite{wheeler:100years,bell:book,aharonov:book}.
The particular difficulties with physical theories that are probabilistic are often aggravated by conceptual issues surrounding probability theory itself, where the profound antagonism between orthodox and Bayesian schools has long clouded a clear view on the subject \cite{jaynes:where}.

Not surprisingly, then, the desire to understand the mathematical framework of quantum theory in terms of simple, more easily comprehensible physical principles is almost as old as quantum theory itself \cite{zeilinger:foundational}.\footnote{
As early as 1928 Hilbert, von Neumann and Nordheim wrote \cite{hilbert:grundlagen}: 
``Bei dieser hohen Bedeutung der Quantenmechanik ist es ein dringendes Bed\"urfnis, ihre Prinzipien so klar und allgemein wie m\"oglich zu erfassen.'' 
(In view of the high significance of quantum mechanics it is an urgent desire to grasp its principles as clearly and generally as possible.) 
}
Attempts at such a principles-based reconstruction of quantum theory are legion.
They have been motivated by a desire to elucidate the conceptual basis and interpretation of quantum theory; 
by the search for (or exclusion of) possible modifications to quantum theory; 
or by the search for alternative mathematical formulations of quantum theory that might be more conducive to, say, a merger with general relativity into a quantum theory of gravity. 
The following are a few examples without any claim to completeness.

\textbf{Quantum logic} is arguably the oldest of the reconstructive approaches.
First put forward by von Neumann in his famous 1932 monograph \cite{neumann:book} and in greater detail together with Birkhoff \cite{birkhoff:logic}, it was subsequently developed and promulgated most prominently by members of the Geneva School \cite{jauch:book,piron:book}.
Its basic idea is to consider a lattice of propositions that is complete, orthocomplemented, weakly modular and atomic.
Such a lattice constitutes a generalisation of classical logic.
The definition of the lattice entails the existence of Boolean ``and'' and ``or'' operations which, however, no longer need to obey classical rules such as distributivity.
According to Piron's theorem \cite{piron:axiomatique} all propositions within such a quantum logic can be identified with subspaces of a Hilbert space over some skew field.

Almost at the same time the \textbf{algebraic approach} originated with the work of Jordan, von Neumann and Wigner \cite{jordan:algebraic}.
It subsequently evolved into the modern theory of $C^*$-algebras \cite{gelfand:embedding,segal:postulates,thirring:book3} which, thanks to its mathematical rigor and prowess, has found successful application in quantum field theory \cite{haag:algebraic,haag:book}.
Rather than a lattice of propositions it takes as its fundamental mathematical object the abstract algebra of operators whose Hermitian elements are identified with physical observables.

A broad class of \textbf{operational approaches} focuses not primarily on the structure of propositions or observables but on primitive laboratory operations such as preparations, reversible transformations and measurements.
Thereby particular emphasis is placed on the probability functionals, or states, and their convex geometry; 
whence this sometimes goes under the name convex set or \textbf{convex cone} approach.
From the convex geometry of the set of states and with the help of additional auxiliary assumptions the full apparatus of quantum theory can be deduced.
Major efforts in this direction have been associated with Mackey \cite{mackey:book}, the Marburg School \cite{ludwig:book,hellwig:operations,kraus:book}, Varadarajan \cite{varadarajan:probability,varadarajan:book}, Gudder \cite{gudder:convex,gudder:book} and others \cite{davies:operational,edwards:operational,araki:statespace,dariano:whatisspecial}.

Branching off this broad current is the somewhat more recent \textbf{test spaces} approach, promoted mainly by members of the Amherst School \cite{foulis:operational1,randall:operational2,foulis:whatare,foulis:filters,foulis:supports,wilce:orthoalgebras}.
It is based on a generalised notion of sample space and emphasises the status of quantum theory as one of many possible generalised probability theories.
Indeed, quantum theory can be regarded as but a variant of classical probability theory:
Like the latter it deals with hypotheses (represented mathematically by subspaces of Hilbert space) and their probabilities (expectation values of the associated projection operators),
with many important theorems of classical probability theory carrying over to the quantum case.
One example is the quantum analog of the de Finetti representation for exchangeable sequences \cite{caves:definettistates}, which in turn is just a special case of a more general result for test spaces \cite{barrett:definettitest}.

Still within the wider context of operational approaches Hardy \cite{hardy:fiveaxioms} recently proposed to derive quantum theory from a simple set of \textbf{``five reasonable axioms''}:
(i) Probabilities are defined as limits of relative frequencies.
(ii) Probability distributions are specified by the minimal number of degrees of freedom that is compatible with the other axioms.
(iii) Systems of the same information-carrying capacity exhibit the same structure, regardless of whether they are isolated or the result of constraining some bigger system.
(iv) Upon combining constituents into a composite system, dimension and number of degrees of freedom are multiplicative.
(v) Between any two pure states there exists a continuous reversible transformation.

The advent of \textbf{quantum computation} \cite{steane:computing,nielsen:book,mermin:book} has led to revived interest in, and cross-fertilisation with, foundational issues \cite{keyl:fundamentals,fuchs:qmasinfo1,barnum:illumination}.
For instance, there have been efforts to distill those computational features that are genuinely quantum mechanical rather than generic to a wider class of non-classical generalised probability theories \cite{barrett:infoprocessing,barnum:cloning,barnum:nobroadcast,barnum:teleportation}.
With quantum theory opening the way to novel forms of computation, some researchers have taken the next logical step to regard quantum theory as \textit{nothing but} a framework for (highly efficient) information processing.\footnote{
This is reminiscent of Bohr's philosophical remark that ``physics is to be regarded not so much as the study of something a priori given, but rather as the development of \textit{methods for ordering and surveying} human experience'' \cite{bohr:book} (my italics).
}
Reconstructive approaches in this spirit include axiomatisations on the basis of category theory \cite{coecke:categorical,harding:categorical} that take very much a computer science perspective, or attempts at characterising quantum theory in terms of purely information-theoretic constraints \cite{clifton:constraints}.

Finally, \textbf{quantum Bayesianism} is the quantum version of the homonymous program in classical probability theory \cite{schack:bayesrule,caves:quantumasbayes,srednicki:probabilities,timpson:bayesianism}.
In the Bayesian view probability theory constitutes an extension of logic \cite{bernardo:book,jaynes:book};
it is but a consistent framework for plausible reasoning in the face of uncertainty.
Probabilities, and hence states, embody some agent's knowledge about, rather than an objective property of, a physical system;
they represent degrees of belief rather than limits of relative frequencies;
and they can be legitimately assigned not just to ensembles but also to individual systems.
Bayesian probability emphasises (and makes explicit) the role of prior knowledge as a key input on a par with measurement data.
Probability assigments are, however, not entirely at an agent's discretion:
They must satisfy a number of consistency requirements which ensure that different ways of using the same information lead to the same conclusions, irrespective of the particular analysis path chosen.
As shown for the classical case by Cox \cite{cox:probability} these consistency requirements manifest themselves mathematically in the sum and Bayes rules.
In the quantum case these are supplemented by additional coherence conditions \cite{fuchs:quantumness}.
There is also an alternative approach that applies Cox-style consistency requirements to amplitudes rather than probabilities \cite{caticha:amplitudes,caticha:insufficient}.

While each of the above approaches captures important aspects of quantum theory, most of them still fall short of singling out quantum theory uniquely or are based on assumptions that call themselves for a more compelling motivation.
For instance, quantum logic does not specify the skew field and cannot exclude the possibility of real \cite{stueckelberg:real} or quaternionic \cite{finkelstein:quaternionic} Hilbert spaces.
Moreover, Piron's theorem holds only for Hilbert space dimension greater than three.
The algebraic approach, though very powerful mathematically, adds little to the conventional formulation of quantum theory in terms of physical understanding.
The convex geometry of the set of states is too weak a constraint to rule out, e.g., generalisations of quantum theory that are nonlinear \cite{mielnik:generalized};
additional assumptions are needed.
For these additional assumptions there are various proposals which, however, have not yet settled on a universally agreed, easily comprehensible set of axioms.
For example, Hardy's five axioms, though intuitively appealing, resort to a questionable ``simplicity'' argument and make strong implicit assumptions about the existence of a tensor product for composite systems.
Information-theoretic constraints can specify the mathematical apparatus of quantum theory only if combined with further assumptions about its algebraic structure.
As for test spaces, it is not obvious how to single out quantum theory from the multitude of possible generalised probability theories;
there are indications, not yet proven, that such would require additional assumptions regarding the symmetry, topology and composition properties of quantum theory \cite{wilce:compactness,wilce:symtop,wilce:topological,wilce:qpl}.
And finally, the Bayesian approach ---when applied to probabilities--- must invoke additional coherence conditions for, e.g., symmetric informationally complete POVM that are of a simple mathematical form but whose full physical meaning is yet to be clarified \cite{appleby:sic};
and when applied to amplitudes (in the form of Cox-style consistency requirements) the question arises why one should start from complex amplitudes in the first place.

In this paper I will not adhere to any specific of the above approaches;
nor do I wish to propose my own reconstructive scheme.
Rather, my more modest goal is to take a step back and have a fresh look at the commonalities and differences between quantum and classical probability.
The purpose of such an analysis is threefold:
to highlight in a comprehensive fashion the (surprisingly many) features that quantum and classical probability have in common;
to distill the (few) properties that truly distinguish them;
and in light of the above, to explore the room there is, if any, for further probabilistic theories other than classical or quantum theory.
In doing so I will strive to keep abstract mathematics to a minimum and instead emphasise as much as possible the physical meaning of the various properties.
A large portion of my paper will be dedicated to the first of the three objectives, as the full range of commonalities of classical and quantum probability is not often exposed --- in contrast to their
fundamental differences, which ever since the Einstein-Bohr debate \cite{epr:complete,bohr:complete} have been scrutinised extensively.
The key (and novel) technical result, on the other hand, will pertain to the second objective:
I will show that the single distinguishing property of quantum theory is the juxtaposition of finite information-carrying capacity and smoothness, where the concept of smoothness will be carefully defined and motivated.
The mathematical derivation of this result will involve close inspection of the symmetry group, with successive constraints leading unequivocally to the unitary group of transformations in complex Hilbert space.
As for the final objective, I will provide arguments why there is likely no further probabilistic theory that satisfies basic physical desiderata.

The outline of this paper is as follows.
In Sections \ref{propositions} through \ref{sec:reductionism} I will identify six areas where classical and quantum theory have important features in common:
the logic of propositions,
symmetry,
probabilities,
composition of systems,
state preparation
and reductionism.
While the two subsections on the logic of propositions and on probabilities are fairly standard and largely inspired by a mix of the existing quantum logic, convex cone and test spaces approaches,
the remaining parts of Section \ref{commonalities} are dedicated to areas that are less often emphasised but in my opinion equally important for the full picture.
Next, in Sections \ref{classical} and \ref{quantum} I will single out joint decidability and smoothness as the distinguishing properties of classical and quantum probability, respectively.
Smoothness in particular will turn out to be a crucial concept, and I will dwell on its definition and physical meaning.
In Section \ref{further} I will explore whether the commonalities identified in Section \ref{commonalities} may constitute an umbrella over not just classical and quantum probability but also other probabilistic theories.
The answer is in principle yes, but the candidate theories will likely have no physical significance.
Finally, in Section \ref{discussion} I will briefly venture outside the umbrella and consider more remote alternatives that share with classical probability less structure than does quantum theory;
but I will quickly dismiss these on physical grounds.
This leads to the conjecture that quantum theory is in fact the only consistent alternative to classical probability.
I will conclude with a brief summary.

\section{\label{commonalities}Commonalities of classical and quantum\\ probability}

\subsection{\label{propositions}Propositions}

\small
\begin{table*}
\begin{tabular}{llll}
 Classical concept&Quantum analog&Generic name&Symbol \\ \hline
 Sample space&Hilbert space&Proposition system&--- \\
 Subset&Subspace&Hypothesis, proposition&$a,b,x,y$ \\
 Element&1-dim. subspace (ray)&Most accurate hypothesis&$e,f$ \\
 Empty set&Zero&Absurd hypothesis&$\emptyset$ \\
 Disjointedness&Orthogonality&Contradiction, exclusion&$\perp$ \\
 Set inclusion&Embedding&Implication, refinement&$\subseteq$ \\
 Set complement&Orthogonal complement&Complement&$\backslash$ \\
 Cardinality&Dimension&Granularity&$d$ \\
 \end{tabular}
\caption{\label{translation}Correspondences between classical and quantum logical structure and their generic representation.}
\end{table*}
\normalsize

Classical and quantum theory share a common logical structure as summarised in Table \ref{translation}.
This common structure is weaker than classical Aristotelian logic in that it lacks the Boolean  ``and'' ($\cap$) and ``or'' ($\cup$) operations, reflecting the fact that in the quantum case hypotheses need not be jointly decidable.
In this weaker structure any proposition $a$ can be endowed with a substructure $({\cal L}_a:=\{x|x\subseteq a\},\subseteq,\backslash)$ isomorphic to an orthomodular poset \cite{birkhoff:book}, with orthocomplementation being defined relative to the maximal element $a$.\footnote{
In the quantum logic approach the ``and'' and ``or'' operations are defined nonetheless as the intersection or closed hull of subspaces, respectively, endowing ${\cal L}_a$ with the richer structure of a lattice rather than of a poset \cite{jauch:book}. 
However, I shall refrain from using these definitions as they violate the classical distributivity property and lack a clear operational meaning.
}

An alternative mathematical description uses instead the partial operation $\oplus$ corresponding classically to the union of disjoint sets, and in the quantum case to the sum of mutually orthogonal subspaces.
This partial sum is symmetric and associative,
\begin{equation}
	x\oplus y=y\oplus x
\end{equation}
\begin{equation}
	(x\oplus y)\oplus z=x\oplus (y\oplus z)=x\oplus y\oplus z
	\quad,
\end{equation}
and has the absurd hypothesis as its unique neutral element
\begin{equation}
	x=\emptyset\quad :\Leftrightarrow \quad\,x\oplus y=y\quad\forall\,y
	\quad.
\end{equation}
By definition a partial sum exists if and only if the summands are mutually contradictory;
or conversely
\begin{equation}
	x\perp y\quad :\Leftrightarrow \quad\exists\,\,x\oplus y
	\quad.
\end{equation}
Logical implication of hypotheses can be defined via
\begin{equation}
	x\subseteq a\quad :\Leftrightarrow \quad\exists\,\,a\backslash x
	\quad,
\end{equation}
where in turn $a\backslash x$ is the complement of $x$ relative to $a$ defined via
\begin{equation}
	y=a\backslash x\quad :\Leftrightarrow \quad\,x\oplus y=a
	\quad.
\end{equation}
The uniqueness of the relative complement renders $({\cal L}_a,\oplus)$ isomorphic to an orthoalgebra \cite{foulis:whatare,foulis:filters,wilce:orthoalgebras} for any choice of maximal element $a$.

Whenever a (classical or quantum) hypothesis $a$ is decomposed into mutually exclusive, collectively exhaustive (MECE) refinements, and each of these refinements into further MECE refinements, and so on in a tree-like fashion until this iterative process comes to a halt because hypotheses cannot be refined any further then regardless of the precise path chosen to arrive at such a maximal decomposition the total number of outermost branches equals the granularity
\begin{equation}
	d(a):=\max \# \{x_i|\bigoplus_i x_i=a, x_i\neq\emptyset\}
	\quad.
\end{equation}
It vanishes if and only if the proposition is absurd;
is equal to one if and only if the proposition is most accurate;
and adds up under partial summation,
\begin{equation}
	d(\bigoplus_i x_i)=\sum_i d(x_i)
	\quad.
\end{equation}
Granularity is closely related to information-carrying capacity:
Having ascertained the truth of hypothesis $a$, the amount of additional information that can be extracted by way of further, more refined measurements is bounded from above ---in both the classical and the quantum case--- by $\log d(a)$.
For simplicity of argument I shall assume in the remainder of this paper that the granularity of all propositions is finite. 

Associated with each hypothesis $a$ are collections of ordered decompositions
\begin{equation}
	{\cal M}_a(k_1,\ldots,k_r):=\{(x_1,\ldots,x_r)|\bigoplus_i x_i=a, d(x_i)=k_i\}
\end{equation}
labelled by the granularity vector $(k_1,\ldots,k_r)$.
These granularities must be non-zero and sum to $d(a)$.
For different granularity vectors the associated collections are mutually disjoint;
while their union, over all $\vec{k}$, contains all possible ordered decompositions of $a$.
Two special cases are the set of all most accurate refinements of $a$,
\begin{equation}
	X_a:=\{e|e\subseteq a\}\sim {\cal M}_a(1,d(a)-1)
	\quad,
\end{equation}
and the collection of unordered maximal decompositions
\begin{equation}
	{\cal A}_a:=\{\{e_i\}|\bigoplus_i e_i=a\}\sim {\cal M}_a(1,\ldots,1)/S_{d(a)}
\end{equation}
which is isomorphic to its ordered counterpart modulo permutation of the branches.
The pair $(X_a,{\cal A}_a)$ constitutes a test space or ``manual'',
which in the test spaces approach is taken as the basic structure on which a probability theory is erected \cite{wilce:orthoalgebras}.

In sum, the common logical structure underlying both classical and quantum probability can be described in either of three equivalent ways:
as a collection of orthomodular posets, of orthoalgebras, or of test spaces.
All these mathematical structures can be associated to arbitrary maximal elements $a$.
In the following I shall limit myself to those cases where the granularity of this maximal element is finite.

\subsection{\label{symmetry}Symmetry}

In both the classical and the quantum case the proposition system can be characterised entirely by its symmetry.
Let ${\cal G}$ be the group of all automorphisms that preserve the partial sum,
\begin{equation}
	g(\bigoplus_i x_i)=\bigoplus_i g(x_i)
	\quad,
\end{equation}
and hence the logical structure exhibited in Table \ref{translation};
and ${\cal G}_a$ its subgroup that leaves in addition the hypothesis $a$ and all its implications invariant,
\begin{equation}
	{\cal G}_a:=\{g\in{\cal G}|g(a\oplus x)=a\oplus x\,\forall\,x\}
	\quad.
\end{equation}
This subgroup also preserves all collections ${\cal M}_a(k_1,\ldots,k_r)$.
Moreover in both the classical and the quantum case it acts on these collections transitively, rendering them homogeneous spaces
\begin{equation}
	{\cal M}_a(k_1,\ldots,k_r)\sim {\cal G}_a / \bigotimes_{i=1}^r {\cal G}_{x_i}\quad,\quad (x_1,\ldots,x_r)\in {\cal M}_a(k_1,\ldots,k_r)
	\quad.
\label{transitivity}
\end{equation}

Collections that pertain to different hypotheses and different granularity vectors are isomorphic whenever the granularity vectors agree up to a permutation,
\begin{equation}
	{\cal M}_a(k_1,\ldots,k_r)\sim {\cal M}_b(l_1,\ldots,l_s)\quad\forall\,\,\{k_i\}=\{l_j\}
	\quad,
\end{equation}
and can thus be grouped into equivalence classes ${\cal M}(\{k_i\})$ labelled by the unordered set of granularities only.
Likewise the subgroups pertaining to different hypotheses fall into equivalence classes that have granularity as their sole parameter.
In effect all structure depends on granularity only, not on any specifics of the proposition system under consideration.\footnote{
That granularity is the sole structural parameter is reminiscent of a conjecture by Fuchs \cite{fuchs:qmasinfo1} that the sole objective property of a quantum system is its Hilbert space dimension.
}
The isomorphism (\ref{transitivity}) then carries over to an isomorphism of equivalence classes
\begin{equation}
	{\cal M}(\{k_i\})\sim {\cal G}(\sum_i k_i) / \bigotimes_i {\cal G}(k_i)
	\quad,
\label{isomorphism}
\end{equation}
including as special cases
\begin{equation}
	X(d)\sim {\cal G}(d)/({\cal G}(d-1)\otimes {\cal G}(1))
\label{isomorphism2}
\end{equation}
and 
\begin{equation}
	{\cal A}(d)\sim ({\cal G}(d)/{\cal G}(1)^{\otimes d})/S_d
	\quad.
\end{equation}
The pair $(X,{\cal A})$ becomes a symmetric ${\cal G}$-test space \cite{wilce:symtop}.

In the classical case ${\cal G}(d)$ is the symmetric (or permutation) group $S_d$;
whereas in the quantum case it is the unitary group $U(d)$.
Whether or not other groups might be physically meaningful is a key issue addressed in this paper.

\subsection{\label{probabilities}Probabilities}

A state $\rho$ assigns to hypotheses a probability between zero and one,
\begin{equation}
	\rho:\,x\to \mbox{prob}(x|\rho)\in[0,1]
	\quad.
\end{equation}
Two states are identical if and only if they yield the same probabilities,
\begin{equation}
	\rho=\sigma \quad\Leftrightarrow\quad \mbox{prob}(x|\rho)=\mbox{prob}(x|\sigma)\,\forall\, x 
	\quad,
\end{equation}
so specifying a state is tantamount to providing a complete list of all probabilities.\footnote{
This definition of the state is operational in the sense that experimental indistinguishability (same probabilities) implies mathematical identification (same states).
}
Classically any probability is given by a sum of elementary probabilities $\sum_{e\subseteq x}\rho(e)$ over the appropriate subset of sample space,
whereas in the quantum case it is given by the trace $\mbox{tr}(\rho P_x)$ (Gleason's theorem \cite{gleason:measures,peres:book}).
In both cases probabilities obey the sum rule
\begin{equation}
	\mbox{prob}(\bigoplus_i x_i|\rho)=\sum_i \mbox{prob}(x_i|\rho)
\label{sumrule}
\end{equation}
as well as the product rule
\begin{equation}
	\mbox{prob}(x|\rho)=\mbox{prob}(x|x\oplus y,\rho)\cdot \mbox{prob}(x\oplus y|\rho)
	\quad,
\label{productrule}
\end{equation}
where the latter is weaker than the classical Bayes rule.
Symmetry transformations of states are defined via the invariance requirement
\begin{equation}
	\mbox{prob}(g(x)|g(\rho))=\mbox{prob}(x|\rho)
	\quad.
\end{equation}

When probabilities are conditional, the order of the conditions may be relevant.
The notation is such that conditions are to be read from right to left:
i.e., $\mbox{prob}(x|y_k,\ldots,y_2,y_1,\rho)$ denotes the probability of $x$ given that one started from the prior $\rho$, then ascertained $y_1$, \textit{subsequently} $y_2$, and so on.
If one of the conditions is a most accurate proposition then ---by the very definition of the term ``most accurate''--- it supersedes all previous conditions.
In particular, it supersedes any prior,
\begin{equation}
	\mbox{prob}(x|e,\rho)=\mbox{prob}(x|e,\sigma)\equiv \mbox{prob}(x|e) \quad\forall\,\,\rho,\sigma
	\quad,
\end{equation}
so that in the posterior probabilities the most accurate proposition effectively plays the role of the new state.
Such a state is termed ``pure''.

Finally, in both the classical and the quantum case the set of states exhibits one further important property:
If $\{\mbox{prob}(x|\rho)\}$ and $\{\mbox{prob}(x|\sigma)\}$ are two lists of probabilities corresponding to states $\rho$ and $\sigma$, respectively, then there exist states yielding any mixture $\{\alpha\mbox{prob}(x|\rho)+\beta\mbox{prob}(x|\sigma)\}$ with $\alpha,\beta\in[0,1]$ and $\alpha+\beta\leq 1$ (not necessarily equal to one as distributions need not be normalised).
States thus form a convex cone. 
In the homonymous approach to quantum reconstruction this geometrical property is taken as the starting point from which, with the help of additional auxiliary assumptions, the full apparatus of quantum theory is deduced.
However, this will not be the approach adopted here.

\subsection{\label{composition}Composition}

Whenever two hypotheses pertain to different physical systems $A$ and $B$ they are jointly decidable\footnote{
This implies a ``no-signalling principle'' \cite{barrett:infoprocessing}.
} 
and can hence be concatenated via the Boolean ``and'' operation
\begin{equation}
	\times:\,x^A, x^B\to x^A\times x^B
	\quad.
\end{equation}
Here the notation ``$\times$'' rather than ``$\cap$'' emphasises the fact that this Boolean ``and'' is not defined in general but only for propositions relating to disparate systems.
Given this restriction, distributivity holds:
\begin{equation}
	(\bigoplus_i x_i^A)\times (\bigoplus_j y_j^B)=\bigoplus_{ij} x_i^A\times y_j^B
	\quad.
\label{distributivity}
\end{equation}
And in both the classical and the quantum case joint probabilities satisfy the product rule
\begin{equation}
	\mbox{prob}(x^A\times y^B|\rho)=\mbox{prob}(x^A|y^B,\rho)\cdot\mbox{prob}(y^B|\rho)
	\quad.
\label{compositeprobability}
\end{equation}

If $e^A$ and $e^B$ are most accurate hypotheses about $A$ and $B$, respectively, then $e^A\times e^B$ constitutes a most accurate hypothesis about the composite $A\times B$.
This implies for the granularity the product rule
\begin{equation}
	d(x^A\times y^B)=d(x^A)\cdot d(x^B)
	\quad.
\end{equation}
Moreover, all propositions of the product form $e^A\times e^B$ are contained in the set of most accurate propositions about the composite system $A\times B$:
\begin{equation}
	X(d_{A}d_{B}) \supseteq X(d_A)\times X(d_B)
	\quad. 
\end{equation}
In the classical case this is in fact an equality because all most accurate propositions about a composite system have the product form.
Consequently, the classical cardinality $\#X(d)=d$ satisfies the product rule $\#X(d_A d_B)=\#X(d_A)\cdot\#X(d_B)$.
In contrast, in the quantum case there exist pure states (and hence most accurate propositions) that are not separable, rendering $X(d_{A}d_{B})$ strictly larger than the product $X(d_A)\times X(d_B)$.
Indeed, this possibility of entanglement is reflected in the manifold dimension $\dim X(d)=2(d-1)$ which for $d\geq 2$ obeys the strict inequality
$\dim X(d_A d_B)>\dim X(d_A)+\dim X(d_B)$.

In a similar vein arbitrary concerted action of reversible operations on different constituents yields an allowed reversible operation on the composite.
In mathematical terms, if ${\cal G}$ is a finite group then the Cartesian product of independent subsets of ${\cal G}(d_A)$ and ${\cal G}(d_B)$ must be isomorphic to an independent subset of ${\cal G}(d_A d_B)$;
or if ${\cal G}$ is continuous, the Cartesian product of Lie generators of ${\cal G}(d_A)$ and ${\cal G}(d_B)$ must be isomorphic to a subset of the Lie generators of ${\cal G}(d_A d_B)$.
This entails the constraint
\begin{equation}
	{\cal G}\left\{
		\begin{array}{ll}
		\mbox{finite:}& \mu'({\cal G}(d_A d_B))\geq \mu'({\cal G}(d_A))\cdot \mu'({\cal G}(d_B)) \\
		\mbox{continuous:}& \dim {\cal G}(d_A d_B) \geq \dim{\cal G}(d_A)\cdot \dim{\cal G}(d_B)
		\end{array}	
	\right.
	\quad,
\label{groupcomposition}
\end{equation}
where in the finite case $\mu'$ denotes the size of the largest independent subset.
Indeed, for the classical symmetric group $S_d$ it is $\mu'(S_d)=d-1$ \cite{whiston:indepsets,cameron:indepsets} and hence the upper inequality is satisfied;
while in the quantum case the lower inequality is satisfied (and even saturated) by the unitary group $U(d)$ with $\dim U(d)=d^2$.

\subsection{\label{preparation}Preparation}

All knowledge about a physical system, embodied in its state, results from a series of experiments or ``preparation procedures''.
Let $a$ denote the proposition that the system under consideration exists at all, and $\sigma$ its state prior to a given procedure.
Each preparation procedure is then an arbitrary combination of
(i) controlled reversible operations;
(ii) measurements\footnote{
In the quantum case the measurements considered here are assumed to be von Neumann measurements rather than POVM.
This does not limit the generality of the argument, as a POVM can always be understood as a von Neumann measurement on some larger system (Neumark's theorem \cite{neumark:spectral,peres:book}).
}
of MECE refinements $\{x_i\}$ of $a$, $\bigoplus_i x_i=a$;
and
(iii) keeping or discarding (= setting $a$ to ``false'') the system, with respective probabilities that may depend both on the prior and on the outcome of the measurement.
As an example consider a photon (so the hypothesis $a$: ``the photon exists'' is true) whose polarisation is first
(i) rotated, then
(ii) measured along some axis with possible outcomes $\{x_1, x_2\}$, $\bigoplus_i x_i=a$, and
(iii) allowed to pass if and only if the outcome is $x_1$, and else discarded (hence $a$ set to ``false'').
These three steps may or may not take place inside a black box hiding the measurement outcome from the observer,
and there may be a finite probability of error.
The net effect on probabilities is then of the general form
\begin{equation}
	\mbox{prob}(y|\sigma)\,\to\,\sum_i \lambda_i(g(\sigma)) \cdot \mbox{prob}(y|x_i,g(\sigma))\equiv \mbox{prob}(y|\rho)
	\quad,
\end{equation}
$g$ being the reversible operation and $\{\lambda_i\}$ the respective probabilities (modulo normalisation) with which the system is kept, given the rotated prior and measurement outcome $x_i$.
The updated probabilities correspond to some posterior $\rho$.
In the special case where the prior is pure ($\sigma=e$) and only a single outcome $x$ is selected, this posterior remains pure ($\rho=f$):
\begin{equation}
	\forall\,e, x{\not\perp} e\,\,\exists \,f\,:\quad \mbox{prob}(y|x,e)=\mbox{prob}(y|f)\,\,\forall\,y
	\quad.
\label{preparationrule}
\end{equation}
The procedure described above can be iterated, with the posterior $\rho$ serving as the new prior for the next iteration, until preparation is completed.
In both the classical and the quantum case all states can be prepared in this way.

An arbitrary probability distribution on the substructure of $a$, i.e., the list of all probabilities $\{\mbox{prob}(x|\rho)|x\subseteq a\}$, is completely specified by a finite number of real parameters.
(One of these parameters is $\mbox{prob}(a|\rho)$ which need not be normalised to one.)
Like the substructure itself the number of parameters depends on the granularity $d$ of $a$ only and shall be denoted by $S(d)$.
Since every distribution results from preparation procedures as described above, an alternative way to specify it is by
(i) the set $\{x_i\}\in {\cal M}(\{k_i\})$, $\sum_i k_i=d$, that was last subjected to measurement, requiring 
$\dim {\cal M}(\{k_i\})$ parameters;
and
(ii) associated with each possible outcome $x_i$, the posterior $\lambda_i(g(\sigma)) \cdot\mbox{prob}(y|x_i,g(\sigma))$ on the substructure of $x_i$, requiring $S(k_i)$ parameters.\footnote{
This presupposes non-contextuality insofar as the posterior on the substructure of $x_i$ and hence the number of parameters $S(k_i)$ do not depend on whichever set of propositions $\{x_j\}_{j\neq i}$ was measured alongside $x_i$.
}
This way of characterising the distribution requires the same total number of parameters, so
\begin{equation}
	S(\sum_i k_i) = \dim {\cal M}(\{k_i\}) + \sum_i S(k_i)
	\quad.
\label{parameterrule}
\end{equation}
Indeed, this condition is satisfied in both the classical and the quantum case.
Classically the set ${\cal M}(\{k_i\})$ is discrete, whence $\dim {\cal M}(\{k_i\})=0$ and $S(d)=d$;
whereas in the quantum case ${\cal M}(\{k_i\})$ is a continuous manifold of dimension
$2\sum_{i<j}k_i k_j$, and $S(d)=d^2$.

\subsection{\label{sec:reductionism}Reductionism}

Let $a^A, a^B$ be hypotheses that pertain to two distinct physical systems $A$ and $B$.
Probability distributions on their substructures are specified by $S(d_A)$ and $S(d_B)$ real parameters, respectively.
Therefore any distribution on one of the substructures can be characterised completely by the probabilities of just some finite set of ---not necessarily mutually exclusive--- hypotheses $\{b_i^A\subseteq a^A\}_{i=1}^{S(d_A)}$ or $\{b_j^B\subseteq a^B\}_{j=1}^{S(d_B)}$, respectively.

Suppose that one knows the $S(d_A)\cdot S(d_B)$ probabilities of the combined hypotheses $\{b_i^A\times b_j^B\}$.
Then these probabilities suffice to specify not only the two single-constituent distributions but also all constituent-constituent correlations.
This can be seen as follows.
First, without loss of generality the hypotheses $\{b\}$ (for either system) can be chosen such that some subset of these, $\{b_i\}_{i\in I}$, $I\subseteq \{1\ldots S(d)\}$ constitutes a MECE decomposition of $a$, $\bigoplus_{i\in I} b_i=a$.
Then distributivity (\ref{distributivity}) and the sum rule (\ref{sumrule}) give all probabilities
\begin{equation}
	\mbox{prob}(a^A\times b_j^B|\rho)=\sum_{i\in I_A} \mbox{prob}(b_i^A\times b_j^B|\rho)
	\quad.
\end{equation}
Next, application of the product rule (\ref{productrule}) yields all
\begin{equation}
	\mbox{prob}(b_i^A|a^A\times b_j^B,\rho)=\mbox{prob}(b_i^A\times b_j^B|\rho) / \mbox{prob}(a^A\times b_j^B|\rho)
\end{equation}
which, as the $\{b_i^A\}$ are informationally complete, implies in fact knowledge of $\mbox{prob}(x^A|a^A\times b_j^B,\rho)$ for any $x^A\subseteq a^A$.
By yet another application of the product rule (\ref{productrule}) one obtains any
\begin{equation}
	\mbox{prob}(x^A\times b_j^B|\rho)=\mbox{prob}(x^A|a^A\times b_j^B,\rho) \cdot \mbox{prob}(a^A\times b_j^B|\rho)
\end{equation}
as well as, via distributivity and sum rule, $\mbox{prob}(x^A\times a^B|\rho)$.
These in turn yield 
\begin{equation}
	\mbox{prob}(b_j^B|x^A\times a^B,\rho)=\mbox{prob}(x^A\times b_j^B|\rho) / \mbox{prob}(x^A\times a^B|\rho)
\end{equation}
and thus, $\{b_j^B\}$ being informationally complete, any $\mbox{prob}(y^B|x^A\times a^B,\rho)$.
Finally, applying the product rule (\ref{productrule}) one last time gives
\begin{equation}
	\mbox{prob}(x^A\times y^B|\rho)=\mbox{prob}(y^B|x^A\times a^B,\rho) \cdot \mbox{prob}(x^A\times a^B|\rho)
\end{equation}
for any $x^A\subseteq a^A, y^B\subseteq a^B$, and hence indeed the complete statistics of both systems including their correlations.

In a reductionist theory the properties of a composite system are determined entirely by those of its constituents and their statistical correlations.
Beyond these there are no genuinely ``holistic'' properties of the composite system.
So knowing the probabilities of $x^A\times y^B$ for any $x^A\subseteq a^A, y^B\subseteq a^B$ is tantamount to knowing the global state of the composite system.\footnote{
This is also known as the ``global state assumption'' \cite{barrett:infoprocessing} or ``local observability principle'' \cite{dariano:whatisspecial}.
}
The $S(d_A)\cdot S(d_B)$ parameters that suffice to specify the former, therefore, also suffice to specify the latter:
\begin{equation}
	S(d_A d_B)\leq S(d_A)\cdot S(d_B)
	\quad.
\label{reductionism}
\end{equation}
This inequality need not necessarily be saturated, as the effective number of degrees of freedom of the composite system might be reduced by constraints on the correlations that are allowed between subsystems.
Classical probability and quantum theory are both reductionist without such constraints, and hence not only satisfy but also saturate the above inequality.

Reductionism is a key prerequisite for the ability to subject probabilistic models to experimental tests, and hence ultimately for the success of the scientific method.
In the modern Bayesian view probabilities have a priori nothing to do with measurable relative frequencies,
a distinction which is particularly apparent in those cases where probabilities pertain to isolated events that cannot be repeated or ---as is the case in quantum theory--- to individual systems that cannot be subjected to measurements without disturbance \cite{hartle:individual}.
Yet whenever probabilities pertain to multi-partite sequences that are exchangeable (or more Bayesian: whose exchangeability is agreed upon by a group of agents) then it \textit{is} possible to collect frequency data, and this data drives agents via Bayes rule towards a unique posterior distribution regardless of their initial, invariably subjective beliefs.
This possibility to reach a consensus through the collection of data is a fundamental assumption underlying any empirical science;
in particular, it is implicit whenever one speaks of ``the'' state of a system as being the result of some well-defined preparation procedure \cite{peres:book,caves:subjective}.\footnote{
Consensus-building may fail, however, if agents do not share the same belief about basic symmetries such as a sequence's exchangeability, and hence start from priors that differ not just in parameter values but also in their parametric form.
It may also fail if for the parameter values agents assign priors that are so different from each other  that they do not have any overlap.
Under such conditions even an infinite amount of measurement data may lead to vastly different conclusions \cite{fuchs:priors}.
Strictly speaking, therefore, the state of a system is never truly objective.
Rather, it constitutes an intersubjective consensus among a group of agents who started from some common set of minimal symmetry assumptions and possibly diverse, yet overlapping priors.
}
But such convergence to a consensus is not self-evident:
It rests on the existence of a de Finetti representation for exchangeable sequences, which in turn is guaranteed only in probabilistic theories that are reductionist \cite{caves:definettistates}.

\section{\label{differences}Additional requirements needed to derive specific cases}

\subsection{\label{classical}Classical case: Joint decidability}

In addition to the shared features discussed in Section \ref{commonalities} classical probability theory makes one more basic assumption:
that all hypotheses be jointly decidable.
Two hypotheses $a$ and $b$ are jointly decidable (denoted $a\leftrightarrow b$) if and only if they have a joint decomposition $\oplus x_i$, 
\begin{equation}
	a\leftrightarrow b\quad:\Leftrightarrow\quad \exists \bigoplus_{i\in I} x_i: \quad
	a=\bigoplus_{j\in I_a\subseteq I} x_j\quad,\quad
	b=\bigoplus_{k\in I_b\subseteq I} x_k
	\quad.
\end{equation}
Under this assumption one can define the Boolean operations ``and'' ($\cap$), ``or'' ($\cup$) for arbitrary propositions via
\begin{equation}
	a\cap b:=\bigoplus_{i\in I_a\cap I_b} x_i\quad,\quad a\cup b:=\bigoplus_{i\in I_a\cup I_b} x_i
\end{equation}
which satisfy the classical distributivity properties
\begin{equation}
	a\cap \bigoplus_j b_j=\bigoplus_j (a\cap b_j)
	\quad,\quad
	a\cap \bigcup_j b_j=\bigcup_j (a\cap b_j)
	\quad.
\end{equation}
The product rule (\ref{productrule}) then implies the classical Bayes rule
\begin{equation}
	\mbox{prob}(a\cap b|\rho)=\mbox{prob}(a|b,\rho)\cdot\mbox{prob}(b|\rho)
\end{equation}
on which, together with the sum rule (\ref{sumrule}), the whole edifice of classical probability theory can be erected \cite{sivia:book,jaynes:book}.

\subsection{\label{quantum}Quantum case: Smoothness}

The distinguishing feature of quantum theory is its peculiar smoothness.\footnote{
There is sometimes the seemingly opposite claim that a characteristic feature of quantum theory is the ``discontinuity'' of state change upon an act of measurement.
Yet in a Bayesian approach such discontinuous changes occur in classical probability theory, too, and simply reflect the process of learning.
}
Despite the limitation that accessible information be finite, quantum theory deals only with continua:
The set of hypotheses about a system, the symmetry group and all probability distributions are continuous.
In mathematical terms, while the information-carrying capacity and hence the granularity $d$ are constrained to be finite the set of most accurate hypotheses $X(d)$ forms a continuous manifold (for $d\geq 2$) of dimension $\dim X(d)>0$,
the simplest example being for qubits ($d=2$) the surface of the Bloch sphere;
this manifold is compact \cite{fivel:interference}.
Correspondingly, the symmetry group ${\cal G}(d)$ is a compact Lie group.
And the continuity of probability distributions manifests itself in the property that given any $a$ and $e_0\in X_a$,
\begin{equation}
	\forall\,\epsilon>0\ ,\ a\supseteq x\supseteq e_0\quad \exists\,\delta>0:
	\ \mbox{prob}(x|e)>1-\epsilon \quad \forall\  e\in {\cal B}_a(e_0;\delta)
\label{continuity}
\end{equation}
where ${\cal B}_a(e_0;\delta)$ is an open ball (in the group-induced topology) in $X_a$ around $e_0$;
i.e., probabilities which are initially equal to one do not suddenly jump to a lower value upon an infinitesimal transformation.
I shall show that this continuity requirement alone, in combination with the commonalities discussed in Section \ref{commonalities}, constrains ${\cal G}(d)$ to be the unitary group $U(d)$ and hence singles out the quantum case.

Before embarking on a proof of this assertion I shall elaborate briefly on the physical meaning of the continuity requirement.
Continuity ensures that probability assignments are robust under small preparation inaccuracies, in a sense which I shall discuss further below.
Moreover, continuity turns out to be linked to the quantum Zeno or ``watched pot'' effect.\footnote{
That quantum theory somehow resolves the tension between discreteness and continuity, and that there may be a link to the Zeno paradox, has been alluded to by Deutsch \cite{deutsch:itfromqubit}.
}
To see this, I first note that being a compact Lie group, ${\cal G}(d)$ is endowed with a positive definite invariant metric \cite{barut:book}.
In this metric let ${\cal G}(d;\delta)$ denote an open ball of radius $\delta$ around the identity.
Continuity then implies that, given any $e\in X(d)$,
\begin{equation}
	\forall\ \epsilon>0\quad \exists\,\delta>0:\ 
	\mbox{prob}(e|g(e))>1-\epsilon\quad\forall\ g\in {\cal G}(d;\delta)
	\quad.
\end{equation}
Define $\delta(\epsilon)$ as the radius which saturates the continuity condition,
\begin{equation}
	\delta(\epsilon):=\sup \{\delta|\mbox{prob}(e|g(e))>1-\epsilon\ \forall\ g\in {\cal G}(d;\delta)\}
	\quad.
\end{equation}
Due to symmetry this radius depends on $\epsilon$ only, not on the specific hypothesis $e$ or granularity $d$.
Then for $N$ replicas
\begin{equation}
	\inf\left.\left\{\prod_{i=1}^N \mbox{prob}(e|g_i(e))\right| g_i\in {\cal G}(d;\delta(\epsilon))\right\}
	=
	(1-\epsilon)^N \approx 1-N\epsilon
	\quad,
\label{1star}
\end{equation}
where the approximation holds whenever $\epsilon$ is sufficiently small.
By composition rule (\ref{compositeprobability}) the $N$-fold product of probabilities can also be written as
\begin{equation}
	\prod_{i=1}^N \mbox{prob}(e|g_i(e))=\mbox{prob}(e^{\times N}|g(e^{\times N}))
\end{equation}
with shorthand $e^{\times N}\equiv e\times\ldots\times e$ and some $N$-partite transformation $g$ taken from the tensor product of open balls ${\cal G}(d;\delta(\epsilon))^{\otimes N}$.
This tensor product is in turn contained in some open ball in the symmetry group ${\cal G}(d^N)$ of the composite system.
The minimal radius of the latter follows from the law of Pythagoras for the group metric;
it equals $\sqrt{N}\delta(\epsilon)$.
The infimum (\ref{1star}) thus translates into
\begin{equation}
	\inf \left.\left\{\mbox{prob}(e^{\times N}| g(e^{\times N}))\right| g\in {\cal G}(d^N;\sqrt{N}\delta(\epsilon))\right\}
	=
	(1-\epsilon)^N\approx 1-N\epsilon
	\ ,
\end{equation}
yielding for sufficiently small $\epsilon$ the scaling property
\begin{equation}
	\delta(N\epsilon)\approx \sqrt{N}\delta(\epsilon)
	\quad.
\label{2star}
\end{equation}
This scaling implies
\begin{equation}
	\mbox{prob}(e^{\times N}|g_1(e)\times\ldots\times g_N(e))\geq 1-\epsilon\quad
	\forall\ g_i\in {\cal G}(d;\delta(\epsilon)/\sqrt{N})
	\quad,
\end{equation}
which can be interpreted as follows:
Given a sequence of $N$ independent and ---to within some finite accuracy--- identically prepared systems,
the composite proposition $e^{\times N}$ is confirmed with asymptotic probability $O(1)$, i.e., larger than an arbitrary threshold $1-\epsilon$ ($0<\epsilon<1$) that does not depend on $N$,
if and only if the constituents were prepared in the pure state $e$ to within an accuracy $O(1/\sqrt{N})$.
In short, continuity ensures that the probabilistic model tolerates finite preparation inaccuracies of the order $O(1/\sqrt{N})$.

A second, related consequence of the scaling property (\ref{2star}) is
\begin{equation}
	\lim_{N\to\infty} \prod_{i=1}^N \mbox{prob}(e|g_i(e))=1\quad
	\forall\ g_i\in {\cal G}(d;\delta/N)
\end{equation}
for arbitrary finite $d$ and $\delta$, a result which is tantamount to the quantum Zeno effect \cite{misra:zeno}:
When the transformation of a pure state $e$ over a finite distance $\delta$ on the group manifold is cut into $N$ steps of equal length,
each followed by a measurement of the original proposition $e$,
then as $N\to\infty$ the net effect is that the system remains trapped in its original state.

I now prove my original assertion that the continuity requirement leads uniquely to quantum theory in complex Hilbert space.
In order to understand  the implications of the continuity requirement for the group ${\cal G}(d)$ I return to its original formulation (\ref{continuity}).
By preparation rule (\ref{preparationrule}), for each $e$ in the open ball ${\cal B}_a(e_0;\delta)$ there exists a unique most accurate hypothesis $f$ such that $\mbox{prob}(f|x,e)=1$;
or equivalently,
\begin{equation}
	\forall\ e\in {\cal B}_a(e_0;\delta)\ ,\ a\supseteq x\supseteq e_0\quad \exists !\, f:
	\quad \mbox{prob}(x\backslash f|x,e)=0
	\quad.
\end{equation}
The product rule (\ref{productrule}) then implies that also
\begin{equation}
	\mbox{prob}(x\backslash f|e)=\mbox{prob}(x\backslash f|x,e)\cdot\mbox{prob}(x|e)=0
	\quad,
\end{equation}
so $e\perp x\backslash f$ and hence
\begin{equation}
	e\subseteq a\backslash (x\backslash f)=:y
	\quad.
\end{equation}
This allows that the overarching hypothesis $a$ can be decomposed in three different ways:
\begin{equation}
	e\oplus (a\backslash e)
	=
	e\oplus (y\backslash e)\oplus (x\backslash f)
	=
	f\oplus (x\backslash f)\oplus (a\backslash x)
	\quad.
\end{equation}
Given $a$ and $x$, and hence $a\backslash x$, one can now specify $e$ in two equivalent ways:
(i) 
directly as $e\in {\cal B}_a(e_0;\delta)$;
or (ii) 
as $e\in X_y$, where in turn $y$ must be specified via $f\in X_x$.
In both cases the total number of parameters needed must be the same:
\begin{equation}
	\dim X(d)=\dim X(d-l+1) + \dim X(l)
	\quad,
\end{equation}
where $d=d(a)$ and $l=d(x)$.
By induction with initial condition $\dim X(1)=\dim X(0)=0$,
\begin{equation}
	\dim X(d)=\dim X(2)\cdot (d-1)
	\quad.
\end{equation}
The isomorphism (\ref{isomorphism2}) together with initial condition $\dim{\cal G}(0)=0$ then constrain the Lie group dimension to be of the quadratic form
\begin{equation}
	\dim{\cal G}(d)=\frac{\dim X(2)}{2}\,d(d-1) + \dim{\cal G}(1)\cdot d
	\quad;
\end{equation}
and for the number of parameters $S(d)$ the dimensional relation (\ref{parameterrule}) together with the initial conditions $S(0)=0$ and $S(1)=1$ yield the similar form
\begin{equation}
	S(d)=\frac{\dim X(2)}{2}\,d(d-1) +  d
	\quad.
\end{equation}

The constraint (\ref{groupcomposition}) on group composition (continuous case) on the one hand, and reductionism (\ref{reductionism}) on the other, allow in fact only a single non-zero value for $\dim X(2)$, namely $\dim X(2)=2$, and only the two values $0$ and $1$ for $\dim {\cal G}(1)$.
These parameter values correspond to the compact Lie groups $\mbox{O}(d)\otimes \mbox{O}(d)$ and $U(d)$, respectively.
Moreover, convexity of the set of states presupposes that $X(d)$ be the hull of a convex set.
Yet for $d=2$ the group $\mbox{O}(2)\otimes \mbox{O}(2)$ yields $X(2)\sim S^1\times S^1$ isomorphic to a torus, which is not simply connected and hence not the hull of a convex set;
similar problems arise with $\mbox{O}(d)\otimes \mbox{O}(d)$ in higher dimensions.
Therefore, when combined with the properties discussed in Section \ref{commonalities}, the continuity requirement leaves indeed only the unitary group $U(d)$ as a permitted symmetry and so leads unambiguously to quantum theory in complex Hilbert space.

\subsection{\label{further}Further cases?}

If neither joint decidability nor smoothness are required then it is conceivable that the commonalities discussed in Section \ref{commonalities} allow for additional cases.
However, I shall argue that the set of possible alternatives to classical or quantum probability is severely limited and likely unphysical.

The isomorphism (\ref{isomorphism}), composition constraint (\ref{groupcomposition}), preparation rule (\ref{parameterrule}) and reductionism (\ref{reductionism}), as well as the initial conditions $\dim {\cal G}(0)=0$, $S(0)=0$ and $S(1)=1$ constrain the number of degrees of freedom to obey a power law
\begin{equation}
	S(d)=d^\mu 
\end{equation}
and the dimension of the group manifold to be of the form
\begin{equation}
	\dim {\cal G}(d)=d^\mu+(\dim{\cal G}(1)-1)\cdot d 
\end{equation}
with some positive integer $\mu>0$ and $\dim{\cal G}(1)\in\{0,1\}$.
The case $\mu=1,\dim{\cal G}(1)=0$ encompasses the classical case ${\cal G}(d)=S_d$ but might in principle also accomodate finite groups other than the symmetric group;
such alternative finite groups might correspond to ``constrained'' versions of classical probability theory.
The case $\mu=1,\dim{\cal G}(1)=1$ yields the same classical sample space but allows for a phase to be attached to each most accurate hypothesis, with associated symmetry group ${\cal G}(d)=U(1)^{\otimes d}$;
this ``semi-classical'' case may be worth exploring further.
The case $\mu=2$ corresponds to the quantum case discussed in Section \ref{quantum} with symmetry group ${\cal G}(d)=U(d)$, possibly modulo some finite factor group.

As genuine non-classical alternatives, therefore, there remain only probabilistic theories of higher order $\mu\geq 3$, a possibility already pointed out by Hardy \cite{hardy:fiveaxioms}.
However, the central result of Section \ref{quantum} implies that such a higher-order theory would necessarily exhibit some form of discontinuity.
Moreover, it appears difficult to imagine the physical meaning of an associated Lie group such as, say, $U(d^2)$ that would be needed to yield the required dimensionalities.
There is of course the possibility that the symmetry group is exceptional or even non-topological;
yet the physical meaning of such an exotic scenario would be even more elusive.
In sum, as long as the features discussed in Section \ref{commonalities} are required to hold,
there appears to be no reasonable alternative to classical or quantum probability.

\section{\label{discussion}Discussion}

The commonalities of classical and quantum probability extend farther than one might at first expect.
As discussed in Section \ref{commonalities}, classical and quantum theory share a substantial number of features regarding the logic of propositions, symmetry, probabilities, composition of systems, state preparation and reductionism.
Only at a late stage do their ways part, with the requirement of joint decidability leading to classical theory,
whereas demanding smoothness leads to quantum theory.
The essence of quantum theory, therefore, is the juxtaposition of finite information-carrying capacity and smoothness:
the fact that even though information-carrying capacity is finite, quantum theory describes systems that are continuous (i.e., subject to a continuum of hypotheses);
or conversely, that even for continuous systems the information-carrying capacity remains finite.
The step from classical to quantum probability thus amounts to ---given finite information-carrying capacity--- forgoing joint decidability in favor of smoothness;
or given smoothness, relinquishing joint decidability in favor of a finite information-carrying capacity.\footnote{
For the purposes of this article I ignore limiting cases where the Hilbert space dimension, and hence information-carrying capacity, may be countably infinite.
}

While from the above perspective the step from classical to quantum probability may now seem rather small, it entails the following well known, far-reaching consequences:

\begin{enumerate}
\item
\textbf{Indeterminism.}
Quantum theory exhibits an irreducible probabilism in the sense that in every state, even if pure, there are always hypotheses whose probabilities are neither $0$ nor $1$.
Mathematically, this manifests itself in non-commutativity and uncertainty relations.
\item
\textbf{Non-separability.}
The whole is more than the sum of its parts; 
it may be in a pure state that is not a product of constituent states.
In contrast to classical separability, the whole can in general not be dissected into parts without a resulting loss of information.
Mathematically, this manifests itself in the possibility of entanglement.\footnote{
This property of quantum mechanics is sometimes interpreted as a peculiar ``holism'' which, however, I find misleading because such terminology suggests a contradiction to the reductionism discussed in Section \ref{sec:reductionism}.
}
\item
\textbf{Observer-dependency.}
Measurement implies disturbance.
The image of reality that emerges through acts of measurement reflects as much the history of intervention as it reflects the external world;
there is no preexisting reality that is merely revealed, rather than influenced, by the act of measurement.\footnote{
An entertaining, albeit loose metaphor for such an innate observer-dependency is Wheeler's modified version of the ``game of 20 questions'' \cite{wheeler:lawwithoutlaw}.
}
Mathematically, this is encapsulated in the Bell and Kochen-Specker theorems \cite{mermin:bell}.
\end{enumerate}

Besides quantum theory there appear to be no other meaningful alternatives to classical probability theory.
I argued in Section \ref{further} that if taken as constraints, the commonalities identified in Section \ref{commonalities} leave little room for cases other than classical or quantum probability.
Of course, those constraints might be relaxed:
At least formally there may exist non-classical theories that share less structure with classical probability than does quantum theory, and that hence may lack some of the features discussed in Section \ref{commonalities}. 
One example is quantum theory in real Hilbert space \cite{stueckelberg:real}.
There the symmetry group is the orthogonal group $\mbox{O}(d)$, and states are specified by $S(d)=d(d+1)/2$ parameters.
Real quantum theory shares with classical probability the first five of the six commonalities discussed in Section \ref{commonalities}.
But it is not reductionist:
Already for a system composed of two real qubits it is $S(4)>S(2)\cdot S(2)$ in violation of Eq. (\ref{reductionism});
there may be ``holistic'' properties of a two-qubit system that cannot be understood in terms of the constituent qubits and their correlations alone. 
As a consequence there is in general no de Finetti representation for exchangeable sequences, which deprives real quantum theory of a crucial link between probabilities and measurable frequencies \cite{caves:definettistates}.

Another non-classical alternative could be quantum theory in quaternionic Hilbert space \cite{finkelstein:quaternionic}.
Its symmetry group is the symplectic group $\mbox{Sp}(d)$ with group dimension $d(2d+1)$.
States are specified by $S(d)=d(2d-1)$ parameters, yielding for two quaternionic qubits the strict inequality $S(4)<S(2)\cdot S(2)$.
This is compatible with reductionism, yet shows that there are constraints on the allowed correlations between subsystems.
All exchangeable sequences have a de Finetti representation but the converse is no longer true:
Not all distributions of the de Finetti form represent allowed states \cite{caves:definettistates}.
The main drawback of quaternionic quantum theory, however, is the fact that it no longer allows
the arbitrary composition of reversible operations.
Already for a system composed of two quaternionic qubits it is $\dim{\cal G}(4)< \dim{\cal G}(2)\cdot \dim{\cal G}(2)$ in violation of the composition constraint (\ref{groupcomposition}).

The above examples strongly suggest that there is in fact no consistent alternative to quantum theory as we know it:
Any non-classical probability theory other than quantum theory in complex Hilbert space likely violates basic physical desiderata.
Future work will be aimed at further corroborating this conjecture.
If this effort is successful then the commonalities exhibited in Section \ref{commonalities} combined with the smoothness requirement of Section \ref{quantum} may provide the basis for a novel reconstruction scheme for quantum theory.

Beyond the further development of reconstruction schemes it is my hope that the results of my investigation may also inform future efforts in other directions: 
e.g.,
the formulation of overarching frameworks for probability theory that encompass both classical and quantum probability as special cases;
the generic definition of notions such as entropy or information without reference to any specific classical, quantum or other representation;
or a deeper conceptual understanding of the essential features of quantum theory that make the genuine difference from classical probability.

\section*{Acknowledgments}

I would like to thank Chris Fuchs, Lucien Hardy, Robert Spekkens and Cozmin Ududec for their hospitality and stimulating discussions during a visit to Perimeter Institute.
I also thank Alex Wilce and Matt Leifer for helpful feedback on an earlier version of this paper.

\newpage

\bibliographystyle{alpha}

\newcommand{\etalchar}[1]{$^{#1}$}

\end{document}